# Spin Hamiltonians in the Modulated Momenta of Light

Juan Feng[1], Zengya Li[1], Luqi Yuan[1], Erez Hasman[2], Bo Wang[1, *], Xianfeng Chen[1,3,4]

[1]*State Key Laboratory of Advanced Optical Communication Systems and Networks, School of Physics and Astronomy, Shanghai Jiao Tong University; Shanghai, 200240, China.*

[2]*Atomic-Scale Photonics Laboratory, Russell Berrie Nanotechnology Institute, and Helen Diller Quantum Center, Technion – Israel Institute of Technology; Haifa, 3200003, Israel.*

[3]*Shanghai Research Center for Quantum Sciences; Shanghai, 201315, China.*

[4]*Collaborative Innovation Center of Light Manipulations and Applications, Shandong Normal University; Jinan, 250358, China.*

*E-mail: wangbo89@sjtu.edu.cn

**Abstract**

Photonic solvers that are able to find the ground states of different spin Hamiltonians can be used to study many interactive physical systems and combinatorial optimization problems. Here, we establish a real-and-momentum space correspondence of spin Hamiltonians by spatial light transport. The real-space spin interaction is determined by modulating the momentum-space flow of light. This principle is formulated as a generalized Plancherel theorem, allowing us to implement a simple optical simulator that can find the ground states for any displacement-dependent spin interactions. Particularly, we use this principle to reveal the exotic magnetic phase diagram from a $J_1$-$J_2$-$J_3$ model, and we also observe the vortex-mediated Berezinskii-Kosterlitz-Thouless dynamics from the XY model. These experiments exhibit high calculation precision by subtly controlling spin interactions from the momentum space of light, offering a promising scheme to explore novel physical effects.

**Introduction**

The collective behavior of numerous natural and social systems—ranging from phase transitions in matter to neural network dynamics and financial market volatility—can be described by a universal spin model that originated in physical science to study magnetism [1]. The Hamiltonian of a spin model is given by $H_r = -\sum_{ij} J_{ij} \mathbf{S}_i \cdot \mathbf{S}_j$, where $J_{ij}$ is the interaction strength between the $i$th and the $j$th spin, with spin **S** being a unit vector, and $J_{ij}$ a function of spin-spin displacement ($\mathbf{r}_{ij}$). The spatial profile of $J(\mathbf{r}_{ij})$ plays a crucial role in determining the system's physical properties. Short-range interactions regulate critical behavior and lead to distinct phase transitions [2], while long-range interactions are likely to induce correlated states for quantum entanglement [3,4]. Particularly, tuning the interaction strength among several neighboring terms in $J(\mathbf{r}_{ij})$ can give rise to exotic magnetic complexity [5,6] or unusual topological effects [7–9]. Therefore, a general $J(\mathbf{r}_{ij})$ has been actively pursued both theoretically and experimentally, for the potential to unravel novel effects and to solve combinatorial optimization problems [10].

Optical platforms offer abundant opportunities for simulating spin models via diverse light-matter interactions, such as nonlinear optical effects [11,12], spontaneous parametric down-conversion [13,14], lasing [15–17], and exciton-polaritons [18,19]. A typical scheme of the optical spin model is supported by an array of resonant structures, such as coupled waveguides [20,21] or microcavities [18,22], where the $J(\mathbf{r}_{ij})$ is realized by the field overlapping between adjacent optical modes through the leakage of evanescent fields. This strategy stands for a group of experimental efforts that aims to construct spin interactions from real-space field overlapping. However, implementing a general $J(\mathbf{r}_{ij})$ necessitates intricate structure design and fabrication, which in turn restricts the functionality and scalability of the spin system.

In addition to real-space approaches, theoretical analysis of spin models can be significantly enriched by examining the momentum spectrum of $J(\mathbf{r}_{ij})$) [23]. For instance, the critical behavior of a short-range interaction corresponds to a second momentum ($\mathbf{k}^2$) in the Fourier transform of $J(\mathbf{r}_{ij})$ [24]. Experimentally, Fourier transform serves as a powerful tool in various spatial light calculations, including image processing [25], spatial differentiation [26–28], and convolution [29,30]. Lately, a spatial photonic Ising machine (SPIM) has been reported as a promising optical architecture for solving large-scale spin models [31]. Leveraging the parallel processing, the optical circuit of the SPIM is well-suited for linear matrix computation with a low power consumption and high operation speed independent of the spin model scalability. The proposed SPIM utilizes the central spot of the momentum space corresponding to a specific all-to-all spin interaction. Based on the primitive version, several upgraded SPIMs have utilized gauge transformation [32] and multiplexing techniques [33,34] to generalize the spin interaction functions.

In this letter, we introduce a momentum-space-modulated spin Hamiltonian, which allows us to implement arbitrary $\mathbf{r}_{ij}$–dependent interactions $J(\mathbf{r}_{ij})$ in a spatial light platform. We establish a general real-and-momentum space correspondence of spin Hamiltonians [35]:

$$H_k = -\sum_{ij} J(\mathbf{r}_{ij})\mathbf{S}_i \cdot \mathbf{S}_j$$
$$= -\iint V(\mathbf{k})I(\mathbf{k})dk_x dk_y. \qquad (1)$$

Here, the sum is taken over all spin pairs including self-interactions. The modulation function $V(\mathbf{k})$ is the Fourier transform of $J(\mathbf{r}_{ij})$, and $I(\mathbf{k})$ is the normalized momentum-space intensity of light. The unit vector $\mathbf{S}$ is mapped to the phase of the optical field $E$ = $\exp(i\varphi)$ via the relationship of $\mathbf{S}$ = $(\cos\varphi, \sin\varphi)$. This results in an XY model for $\varphi \in (-\pi, \pi]$, and an Ising model for $\varphi \in \{0, \pi\}$. The relationship established in Eq. (1) reveals a general correspondence between a two-body spin Hamiltonian and a $V(\mathbf{k})$-modulated diffraction of light, allowing us to simulate spin model behaviors via optical momentum modulations. Specifically, for $V(\mathbf{k}) = 1$, Eq. (1) reduces to the Plancherel theorem. For $V(\mathbf{k}) = \delta(\mathbf{k})$, it maps to a specific all-to-all spin interaction. Therefore, short-range interactions, long-range interactions, and arbitrary $J(\mathbf{r}_{ij})$ can be feasibly implemented by applying different $V(\mathbf{k})$. As examples, we perform two groups of experiments to show the ability of our optical simulator. The first experiment applies a $J_1$-$J_2$-$J_3$ Ising spin model by tuning the interaction strength ratio among the nearest neighbor (NN), next-to-NN, and 3$^{rd}$ NN terms, demonstrating distinct magnetic ground states that occur in iron chalcogenides. The second experiment performs optical annealing for the XY spin model with NN interaction, exhibiting a Berezinskii-Kosterlitz-Thouless (BKT) dynamics that is governed by vortex proliferation. These experiments exhibit high calculation precision, with application potentials to solve general spin models using a simple spatial light architecture.

**Main text**

*The working principle of optical spin model simulator.* As illustrated in FIG. 1, a plane wave laser impinges onto a phase-only spatial light modulator (SLM), generating a square array of optical fields with uniform amplitude and arbitrary phases. The phase of light is mapped to the spin vector in the same way as we have mentioned before (FIG.1 (a) and (b)). The lens transforms the real-space light into the Fourier spectrum $I(\mathbf{k})$, which is captured by a camera [35]. The optical simulation of finding the ground state is accomplished by cooperating the optical system with a computer, which generates random phases via a Markov chain [36] and accepts new phase configurations according to the Metropolis algorithm [37]. This iterative process ensures that light evolves towards smaller $H_k$.

*$J_1$-$J_2$-$J_3$ model experiments.* The $J_1$-$J_2$-$J_3$ model plays a pivotal role in revealing the intricate magnetism complexity of iron chalcogenides [5,38,39]. The Hamiltonian of $J_1$-$J_2$-$J_3$ model is given by $H_r = -\sum_{NN} J_1 \mathbf{S}_i \cdot \mathbf{S}_j - \sum_{2NN} J_2 \mathbf{S}_i \cdot \mathbf{S}_j - \sum_{3NN} J_3 \mathbf{S}_i \cdot \mathbf{S}_j$. Here, the first sum is taken over all NN spin pairs, the second sum over next-to-NN spin pairs, and so on. For simplicity, we set $J_1 = 1$, and we define two parameters to describe the interaction ratios, $R_1 = J_2/J_1$, and $R_2 = J_3/J_1$, akin to the notation in Ref. [6]. The theoretically calculated phase diagram for the $J_1$-$J_2$-$J_3$ model at $T = 0$ is depicted in Fig. 2 (Ref. [6]). Here, $T$ is the temperature normalized by $J_1/k_B$. There are four types of states separated by linear boundaries determined by the interaction ratios $R_1$ and $R_2$.

Among them, the antiferromagnetic ground state is the most common one that can be achieved without considering $J_2$ and $J_3$. As $R_1$ and $R_2$ increase, distinct phases emerge, such as the double-stripe pattern (4×4) and staggered-dimer pattern (4×2) (FIG. 2). These patterns are important ground state candidates for the iron chalcogenides.

We apply the $J_1$-$J_2$-$J_3$ model to our simulator to solve the ground states, starting with the case of $(R_1, R_2) = (0.5, 0.9)$. The interaction function $J(\mathbf{r}_{ij})$ and the Fourier transform $V(\mathbf{k})$ are calculated and shown in FIG. 3(a). A 10×10 random spin configuration is initially encoded onto the SLM, as depicted in FIG. 3(b). Meanwhile, the diffraction of light in the momentum space is captured by the camera. Because of the binary phase encoding for the Ising model, the diffraction in the momentum space is centrosymmetric with $I(\mathbf{k}) = I(-\mathbf{k})$. In order to reach the ground state without being trapped in a local minimal, we have divided the annealing process by 10 temperatures from $T = 1.8$ to $T = 0$ (FIG.3(e)). At every temperature, we run 2000 iterative steps in order to update the phase configuration to reach thermal equilibrium [35]. During this process, the experimentally obtained $H_k$ is depicted in FIG.3(e). For high-$T$ cases, the $H_k$ is overall decreasing as the iteration continues, but it also exhibits strong fluctuation due to the simulated thermal effect from the Metropolis algorithm. At around $T \sim 1$, a minimal $H_k$ is reached in good agreement with the theoretically predicted critical temperature FIG. 3(f). The annealing process is terminated after $2 \times 10^4$ iterative steps and a double-stripe pattern (4×4) is observed in FIG. 3(c), which agrees with the theoretical prediction in FIG. 2(b). The momentum space intensity $I(\mathbf{k})_{sol.}$ for the ground state is shown in the lower panel of FIG. 3(c), where we see four diffraction spots. During the experiment, a linear relationship between $H_k$ and $H_r$ is observed to verify Eq. (1), as depicted in FIG. 3(d).

Next, we systematically vary the interaction ratios $R_1$ and $R_2$ to uncover the complete phase diagram in FIG.2. Particularly, the AF state (Fig. 4(a)), single-stripe (FIG. 4(d)), and double-stripe (FIG. 4(c)) are all in perfect agreements with the theoretical predictions. For $(R_1, R_2) = (0.5, 0.3)$, the solved state is slightly higher than the predicted ground state due to the competitive interaction of the staggered-stripe pattern (FIG. 4(b)). The newly observed state for $(R_1, R_2) = (0.5, 0.9)$ is a mixture of two cases, which is different from that in FIG. 3(c). This phenomenon arises because of the ground state degeneracy since random flips during optical annealing can stochastically choose one spin distribution over another, leading to a different or mixture of ground state configurations. From the experimental results we can see that as $V(\mathbf{k})$ modulates the flow of light, it does not fix the spatial profile of the final $I(\mathbf{k})$. Therefore, there is no target image as proposed in ref. [31], since the Hamiltonian is defined by a real-valued number from Eq. (1), rather than an intensity distribution. More importantly, a single target image cannot map to all ground states of a spin system, especially for a frustrated spin system that has many degenerated modes.

*BKT dynamics*. We extend the use of our simulator to solve an XY model by extending the spins as quasi-continuous variables. A well-known phenomenon from the XY model is the BKT phase transition, which is a critical phenomenon characterized by the number of vortices [40]. For simplicity, we consider a ferromagnetic NN interaction that is reduced from the $J_1$-$J_2$-$J_3$ model by setting $J_1 = -1$ and $J_2 = J_3 = 0$. Practically, a SLM can only generate discrete phases. Therefore, we utilize a $q$-state clock model [41,42] to approach the XY model by dividing a continuous $2\pi$ phase into $q$ levels. In our experiments, we encode an array (20×20) of random phases onto the SLM with $q = 8$. Typical phase distributions at different temperatures are shown in FIG. 5(a). It can be seen that the high-$T$ system is strongly affected by thermal noise, and the phases are changing abruptly between neighboring spins. When the system is gradually cooling down, random phases tend to be collinear with each other by the NN spin interaction. The topological phase transition is characterized by the evolution of vortex number ($N_v$) as a function of $T$, as depicted in FIG. 5(b). The $N_v$ crossover occurs around $T = 1$, agreeing with the numerical calculation (critical temperature $T_c \sim 0.9$). Meanwhile, we capture $H_r$ as a function of $T$, as presented in Fig. 5(c). The Hamiltonian of the XY system remains at a temperature of $T \approx 0.5$, indicating the optical noise level of our system.

At last, we perform a quenching experiment to showcase the out-of-equilibrium dynamics of the XY spin system. To do that, we start with a random phase distribution and set $T=0$. The quenching dynamics evolve through a coarsening process revealed in the $H_k$ curve presented in FIG.5 (e) [35,43]. The final spin configuration is trapped in a typical vortex-pair state as shown in FIG. 5(c).

*Conclusion*. In summary, we have demonstrated an optical simulator to find the ground states of distinct spin models by exploiting the momentum-space modulation of light. This approach opens the avenue to manipulating the spin interaction from a conjugate space, which widely exists in many classical and quantum systems [44–46]. In this work, we have utilized linear optics to demonstrate two important physical phenomena, including the complex ground states from the $J_1$-$J_2$-$J_3$ model, and the BKT transition from the XY model. It has been shown that a quadratic spin interaction can be realized using nonlinear optical effects [47,48], which can be combined with our moment-space modulation method to implement a $J_1$-$J_2$-$J_3$-$K$ model [5] to show the magnetism complexity of iron-based superconductors. Moreover, the implementation of an XY model is a critical step to generalize optical simulators with continuous spin variables adapting a wider group of combinatorial optimization problems. Leveraging the versatile multiplexing technique of optics, our system can be utilized to realize asymmetric or nonreciprocal interactions, such as the Dzyaloshinskii–Moriya [49,50] and Haldane interactions to simulate nontrivial topological effects. Moreover, the architecture of our spin model simulator is compatible with intracavity optical systems [51–53]. In this sense, $V(\mathbf{k})$ can be performed by SLMs or high-performance metasurfaces [54–57] that are placed in the momentum plane of an intracavity system with coupled lasers or exciton-polaritons. Optimization algorithms [58] and electric circuits [59] can also be applied to enhance the computation performance of our simulator for better application purposes.

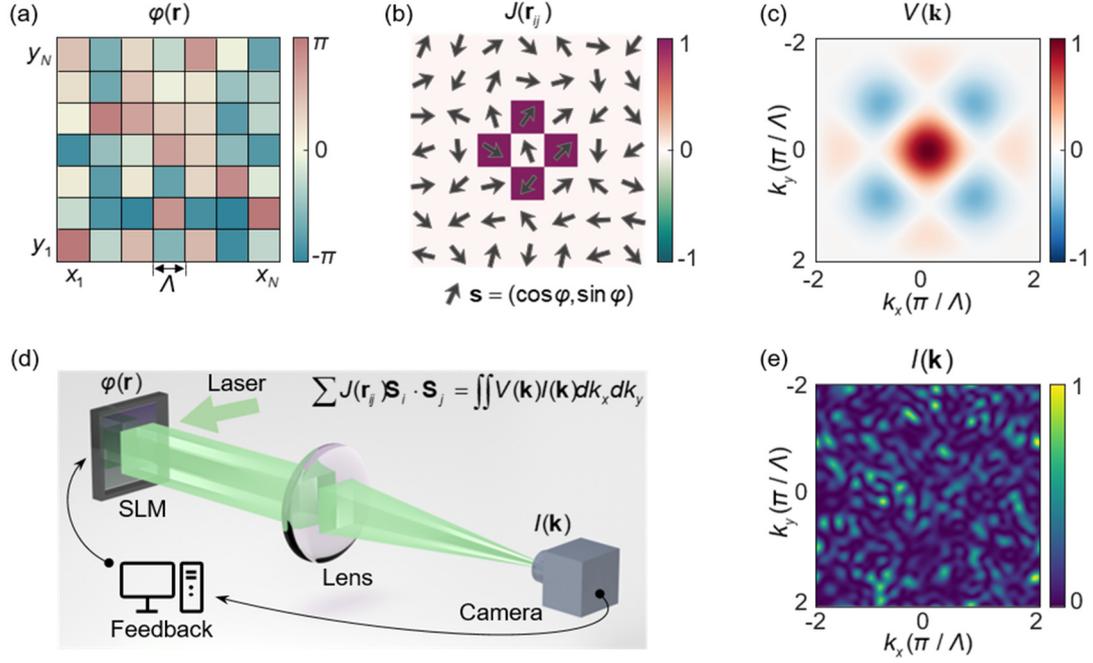

**FIG. 1 Schematic of the optical spin model simulator.** (a) The spin array is represented as the phases of light. The spin lattice constant is $\Lambda$, and the size of the array is $N \times N$. (b) An example of the spin interaction function with only the NN term. The black arrows represent the XY spin vectors mapped to the phases of light. (c) The Fourier transform $V(\mathbf{k})$ of NN interaction function. (d) Simplified experimental setup. A plane wave laser beam (wavelength of 532nm) is phase-modulated via a reflective SLM, collected by a lens, and detected by a CMOS camera in the momentum space (Fourier plane). Feedback is implemented by a computer for data processing, calculating the Hamiltonian, and updating the phase distributions. (e) An example of light momentum-space intensity distribution captured by the camera.

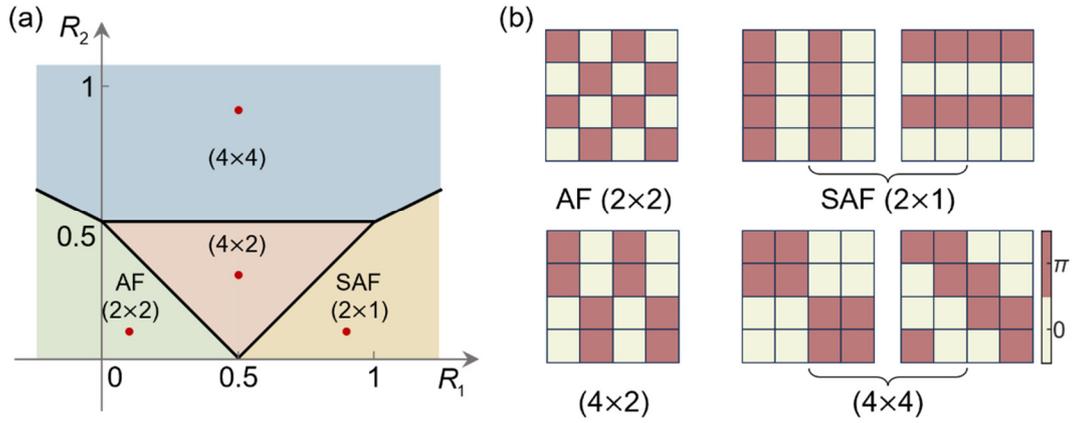

**FIG. 2 Phase diagram of a $J_1$-$J_2$-$J_3$ model at $T = 0$.** (a) The predicted phase diagram at T = 0 as a function of $R_1$ and $R_2$. The red dots denote the locations where we conduct experimental demonstrations. (b) The ground states correspond to the four regions in (a). AF: antiferromagnetic state; SAF: super-antiferromagnetic state; The notation (2×2) means that the unit cell of the ground state is composed of a 2×2 spin array. The state (4×2) is also known as the staggered-stripe pattern, (2×1) is the single-stripe pattern, and (4×4)s are the double-stripe patterns.

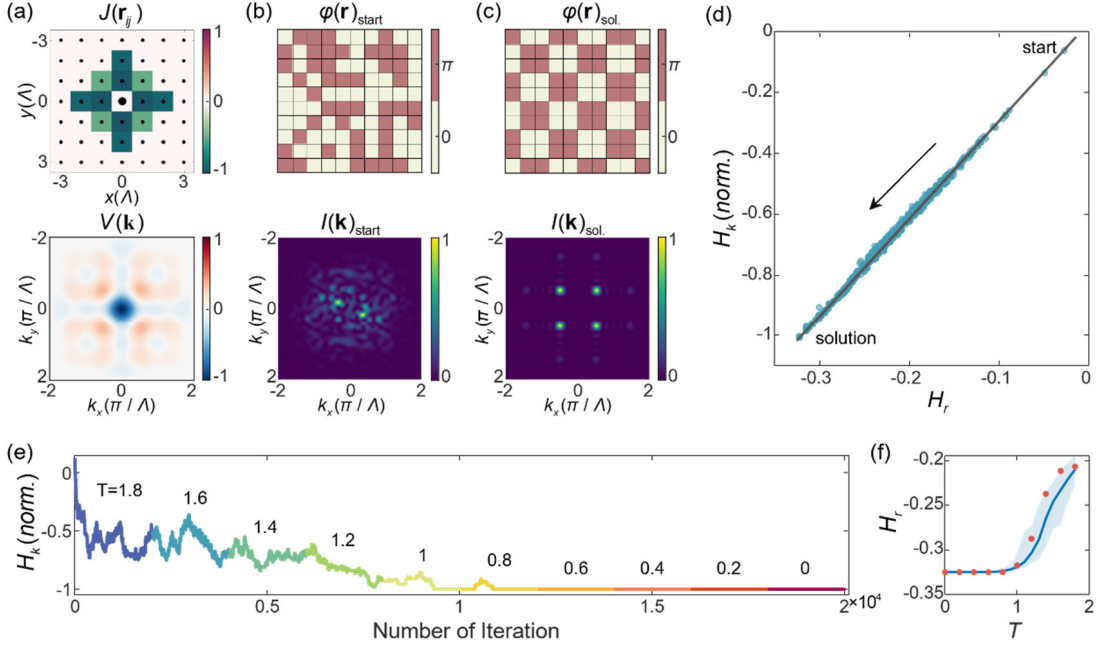

**FIG. 3 Optically solving a typical ground state of the $J_1$-$J_2$-$J_3$ model**. (a) The spin-interaction function $J(\mathbf{r}_{ij})$ for $(R_1, R_2) = (0.5, 0.9)$ and the corresponding Fourier transform $V(\mathbf{k})$. The dots array stands for the locations of spins. (b) and (c) are the phase distributions (upper panel) and corresponding momentum-space intensity distributions (lower panel) for an initial spin distribution (b) and the solved spin configuration (c). (d) The recorded $H_r$-$H_k$ during the optical annealing (dots). The black line is a fitted result, and the black arrow indicates the evolution direction. (e) The observed $H_k$ during optical annealing. (f) $H_r$ as a function of $T$. The dots are experimental results, the curve is the averaged simulation result, and the shaded area is the simulation variance from statistics. Note that $H_r$ is normalized by the number of spin-spin interactions, and we have set the Boltzmann constant as $k_B = 1$.

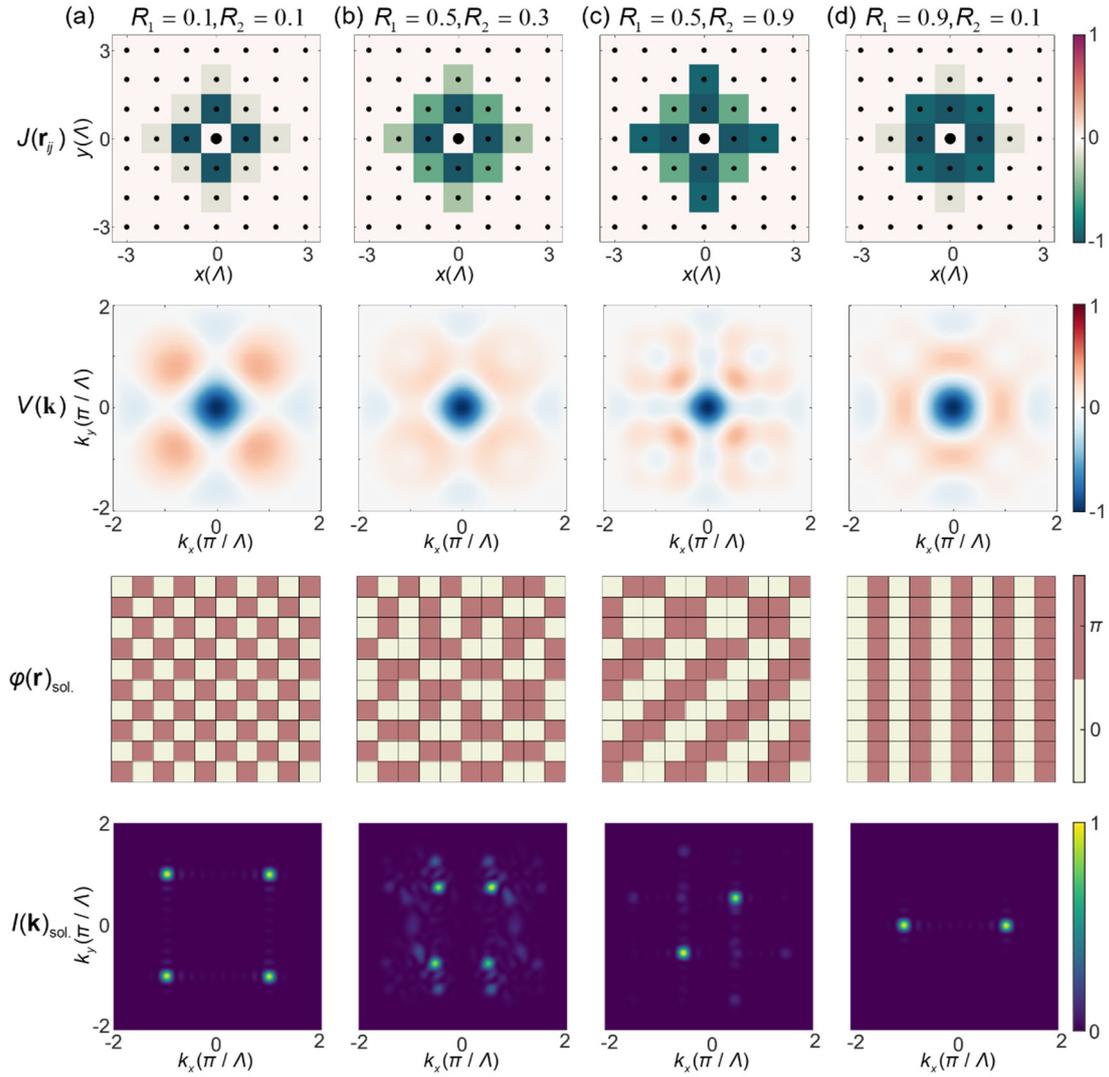

**FIG.4 Experimental demonstrations for the $J_1$-$J_2$-$J_3$ model phase diagram with optical momentum modulations.** From the top to the bottom panels: the spin-interaction function $J(\mathbf{r}_{ij})$, the Fourier transform $V(\mathbf{k})$, the solved ground states $\varphi(\mathbf{r})_{sol.}$, and the corresponding diffraction image $I(\mathbf{k})_{sol.}$.

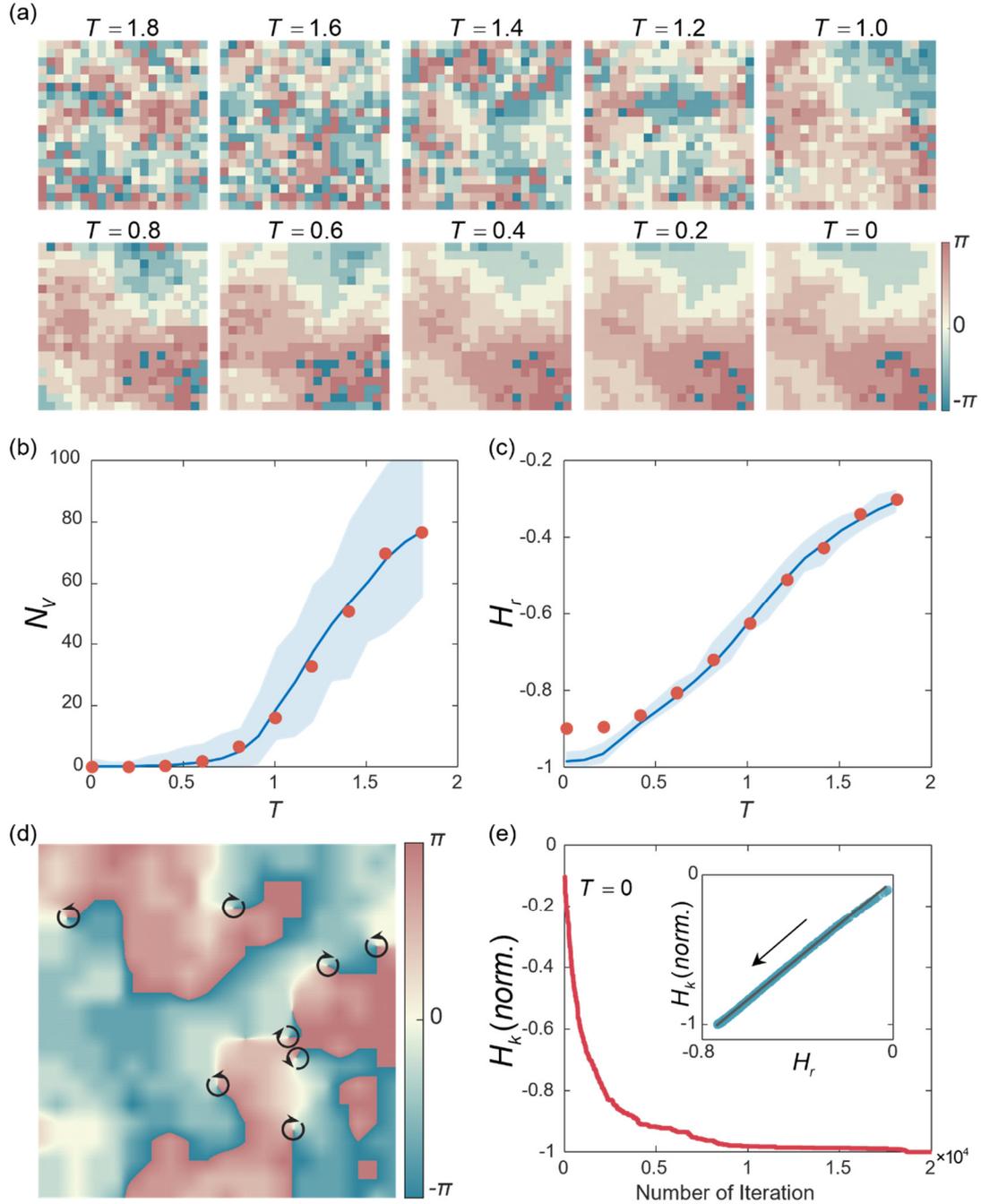

**FIG. 5 Observation of BKT dynamics.** (a) Observed spin distributions at different temperatures. (b) The number of vortices as a function of $T$. The dots are experimentally extracted from (a), and the curve is an averaged simulation result, with a statistical variance denoted as the shaded area. (c) Observed $H_r$ as a function of $T$. (d) The observed vortex-pair state from quenching. The curved black arrows indicate the opposite signs of topological charges. (e) The experimentally recorded $H_k$ during quenching. The inset shows the observed $H_k$ - $H_r$ correspondence.

*Acknowledgment*. We gratefully acknowledge the financial support from the National Key Research and Development Program of China (2022YFA1205100), National Science Foundation of China (12274296, 12122407, 12192252), Shanghai International Cooperation Program for Science and Technology (22520714300). B.W. and L.Y are sponsored by the Yangyang Development Fund. E.H. gratefully acknowledges financial support from the Israel Science Foundation (grant number 1170/20)